\documentclass[12pt]{article}

\renewcommand{\d}{{\rm d}}

\newcommand{\overOm}{\overline{\Omega}}
\newcommand{\overT}{\overline{T}}
\newcommand{\overR}{\overline{R}}
\newcommand{\overU}{\overline{U}}
\newcommand{\overtilR}{\overline{\tilde{R}}}
\newcommand{\dotOm}{\dot{\Omega}}
\newcommand{\dotom}{\dot{\omega}}

\begin{document}

\title{First order discrete Faddeev gravity at the strongly varying fields
}

\author{V.M. Khatsymovsky \\
 {\em Budker Institute of Nuclear Physics} \\ {\em of Siberian Branch Russian Academy of Sciences} \\ {\em
 Novosibirsk,
 630090,
 Russia}
\\ {\em E-mail address: khatsym@gmail.com}}
\date{}
\maketitle

\begin{abstract}
We consider the Faddeev formulation of general relativity (GR), which can be characterized by a kind of $d$-dimensional tetrad (typically $d$=10) and a non-Riemannian connection. This theory is invariant w. r. t. the global, but not local, rotations in the $d$-dimensional space. There can be configurations with a smooth or flat metric, but with the tetrad that changes abruptly at small distances, a kind of "antiferromagnetic" structure.

Previously, we discussed a first order representation for the Faddeev gravity, which uses the orthogonal connection in the $d$-dimensional space as an independent variable. Using the discrete form of this formulation, we considered the spectrum of (elementary) area. This spectrum turns out to be physically reasonable just on a classical background with large connection like rotations by $\pi$, that is, with such an "antiferromagnetic" structure.

In the discrete first order Faddeev gravity, we consider such a structure with periodic cells and large connection and strongly changing tetrad field inside the cell. We show that this system in the continuum limit reduces to a generalization of the Faddeev system. The action is a sum of related actions of the Faddeev type and is still reduced to the GR action.
\end{abstract}

keywords: Einstein theory of gravity; composite metric; Faddeev gravity; minisuperspace model; lattice gravity; piecewise flat spacetime; connection


MSC classes: 83C99; 53C05

\section{Introduction}

The formulation of GR proposed by Faddeev \cite{Fad} considers the metric tensor as a composite field $g_{\lambda\mu} = f^A_\lambda f_{\mu A}$, a function of $d$ vector fields, $f^A_\lambda$, $A$ = 1, \dots, $d$. These fields can be viewed as a $d$-dimensional tetrad as well. For the convenience of notation, we use the Euclidean signature of the metric; the corresponding $d$-dimensional space is assumed to be Euclidean. The formulation of interest is obtained by introducing some metric-compatible affine connection with torsion,
\begin{equation}                                                            
\tilde{\Gamma}^\lambda_{\mu \nu} = f^\lambda_A f^A_{\mu , \nu},
\end{equation}

\noindent alternative to $\Gamma^\lambda_{\mu\nu}$, and using it in the Riemann tensor. The action $\int R \sqrt{g} \d^4 x$ is
\begin{equation}\label{Faddeev}                                             
\int \left( f^\lambda_{A , \lambda } f^\mu_{A , \mu } - f^\lambda_{A , \mu } f^\mu_{A , \lambda } \right) \Pi^{AB} \sqrt{g} \d^4 x, ~~~ \Pi^{AB} = \delta^{AB} - f^{\lambda A} f^B_\lambda,
\end{equation}

\noindent instead of the Hilbert-Einstein one. Here $\Pi^{AB}$ is a projector which projects orthogonally onto the subspace spanned by the tetrad. It was called the {\it vertical} projector.

To show that (\ref{Faddeev}) is equivalent to the Hilbert-Palatini action, we project the equations of motion for $f^\lambda_A$ vertically,
\begin{equation}\label{vertical}                                            
\delta S \equiv \int \frac{\delta S}{\delta f^\lambda_A} \delta f^\lambda_A \d^4 x, ~~~ 0 = \Pi_{AB} \frac{\delta S}{\delta f^\lambda_B} \frac{1}{2 \sqrt{g}} = b^\mu_{\mu A} T^\nu_{\nu \lambda} + b^\mu_{\lambda A} T^\nu_{\mu \nu} + b^\mu_{\nu A} T^\nu_{\lambda \mu}.
\end{equation}

\noindent Here $b^\lambda_{\mu A} = \Pi_{AB} f^{\lambda B}_{, \mu}$, $T^\lambda_{\mu \nu} = \tilde{\Gamma}^\lambda_{\mu\nu} - \tilde{\Gamma}^\lambda_{\nu \mu}$. These can be considered as equations for torsion $T^\lambda_{\mu \nu}$. If $d = 10$, there are effectively $d - 4 = 6$ values of the index $A$ (in the orthogonal subspace) and therefore $6 \times 4 = 24$ vertical equations (\ref{vertical}) just for the 24 components $T^\lambda_{\mu \nu}$. The determinant of this system turns out to be nonzero almost everywhere, and we assume that the system (\ref{vertical}) is equivalent to the vanishing of torsion $T^\lambda_{\mu \nu}$. Then $\tilde{\Gamma}^\lambda_{\mu \nu} = \Gamma^\lambda_{\mu \nu}$, and the action (\ref{Faddeev}) is in fact the Hilbert-Einstein one.

The Faddeev gravity can be attributed to some class of the gravity theories, where the metric is a secondary notion derived from certain more fundamental fields, as, for example, in \cite{Chiu}.

A specific feature of the Faddeev gravity is that the action is finite even if $f^A_\lambda (x )$ and therefore $g_{\lambda \mu} (x )$ are discontinuous (stepwise). This is because the action does not contain any of the squares of derivatives. This makes it possible to partition the spacetime into regions that do not virtually coincide geometrically on their common boundaries and can be considered virtually independent. In particular, any surface in quantum theory can be considered as consisting of independent (although, of course, interacting) areas, each of which can have its own spectrum. The area spectrum plays an important role in the black hole physics.

As usual, the Hamiltonian formalism, useful for studying the spectra of canonical variables, is most easily constructed proceeding from the first order gravity. The formalism of the second order was just considered, and we also proposed a formalism of the first order. In analogy with the usual so(3,1) connection representation of GR, now one would expect the Cartan-Weyl form with so(10) connection. Further, the Faddeev action is invariant w. r. t. the global SO(10) rotations, but not the local ones. Correspondingly, the representation of interest includes a term which violates the local SO(10) symmetry. Besides that, the parity violating term typical for the connection formalism can be present here as well with an analog $\gamma_{\rm F}$ of the Barbero-Immirzi parameter $\gamma$ \cite{Barb,Imm}. It is a coefficient at a term which can be added in the connection representation of the GR action without changing the result of excluding this connection via equations of motion \cite{Holst,Fat}. This leads to some natural generalization of the genuine Faddeev action by including a parity violating term. This generalization is still equivalent to GR. The considered representation takes the form \cite{Kha1}
\begin{eqnarray}\label{S Imm full}                                          
& & \hspace{-0mm} S = \int \left ( f^\lambda_A f^\mu_B + \frac{1}{2 \gamma_{\rm F} \sqrt{g}} \epsilon^{\lambda \mu \nu \rho} f_{\nu A} f_{\rho B} \right ) \left [ \partial_\lambda \omega^{AB}_\mu - \partial_\mu \omega^{AB}_\lambda \right. \nonumber \\ & & \left. + (\omega_\lambda \omega_\mu - \omega_\mu \omega_\lambda)^{AB} \right ] (\omega ) \sqrt{g} \d^4 x + \int \Lambda^\nu_{\lambda \mu } \omega_{\nu}^{AB} \left ( f^\lambda_A f^\mu_B - f^\mu_A f^\lambda_B \right ) \sqrt{g} \d^4 x.
\end{eqnarray}

\noindent Here, $\Lambda^\nu_{\lambda \mu }$ are the Lagrange multipliers at the (violating local SO (10) symmetry) constraint, which expresses the vanishing of the horizontal-horizontal block in $\omega$.

The canonically conjugate variables are some tetrad bilinears and some components of so(10) connection. The so(10) (infinitesimal) connection can vary in infinite limits, and the tetrad bilinears possess continuous spectrum, but have an indirect relationship with the surface area.

In the discrete version, the notion of elementary area appears, and finding its spectrum acquires a sense. The spectrum of area of any surface will be the sum of spectra of elementary areas. The canonically conjugate variables are bivectors of area themselves and some components of the discrete connection. The latter are finite rotations in the $d$-dimensional space-time, and the conjugate area can have a discrete spectrum.

The discrete Faddeev gravity can be constructed on the piecewise flat or simplicial spacetime, like the Regge calculus \cite{Regge} (see also review \cite{RegWil}). In Regge calculus, the metric is approximated by the piecewise flat one, or the metric on the collection of the flat 4-dimensional tetrahedra or 4-simplices \cite{piecewiseflat=simplicial}. The Causal Dynamical Triangulations approach \cite{cdt} related to the Regge calculus proves to be effective in numerical quantum gravity simulations.

A diffeomorphism invariant discrete analogue of the continuum field $f^A_\lambda$ on the piecewise constant ansatz is $f^A_\lambda \Delta x^\lambda_{\sigma^1} \equiv f^A_{\sigma^1}$ for the edges (1-simplices) $\sigma^1$. The connection is SO(10) matrix $\Omega^A_{\sigma^3 B}$, a function on the tetrahedra (3-simplices) $\sigma^3$. The possibility for these fields $f^A_\lambda$ in any neighboring 4-simplices to be independent allows us to use a combinatorially simpler decomposition of space-time into 4-dimensional cubes to approximate any field $f^A_\lambda (x )$. Then we can associate with any vertex a certain four edges $\sigma^1_i, i = 0, 1, 2, 3,$ that form some 4-cube $\sigma^4$. It is convenient to take integer values of the coordinates of the vertices so that $\Delta x^\lambda_{\sigma^1_i} = \delta^\lambda_i$ and formally our discrete variables $f^A_{\sigma^1_i}$ in the 4-cube $\sigma^4$ formed by $\sigma^1_i, i = 0, 1, 2, 3,$ are the values of $f^A_i (x )$ at the vertices of the lattice. The connection variable $\omega_\lambda \Delta x^\lambda_{\sigma^1_i} = \omega_i$, the generator of $\Omega_{\sigma^3}$ on the 3-face $\sigma^3$ formed by $\sigma^1_k, k \neq i$.

In these variables, the discrete first order Faddeev action takes the form \cite{Kha2}
\begin{eqnarray}\label{S-discr}                                             
& & S^{\rm discr} = 2 \sum_{\rm sites} \sum_{\lambda, \mu} \left[ \sqrt{v^{\lambda \mu} \circ v^{\lambda \mu} } \arcsin \frac{v^{\lambda \mu} \circ R_{\lambda \mu}(\Omega ) }{\sqrt{v^{\lambda \mu} \circ v^{\lambda \mu} } } \right. \nonumber \\ & & \left. + \frac{1}{\gamma_{\rm F}} \sqrt{V^{\lambda \mu} \circ V^{\lambda \mu} } \arcsin \frac{V^{\lambda \mu} \circ R_{\lambda \mu}(\Omega ) }{\sqrt{V^{\lambda \mu} \circ V^{\lambda \mu} } } \right]  + \sum_{\rm sites} \sum_{\lambda, \mu, \nu} \Lambda^\nu_{\lambda \mu} \Omega^{AB}_\nu (f^\lambda_A f^\mu_B - f^\mu_A f^\lambda_B), \nonumber \\ & & R_{\lambda\mu} (\Omega ) = \overOm_\lambda (\overT_\lambda \overOm_\mu) (\overT_\mu \Omega_\lambda) \Omega_\mu .
\end{eqnarray}

\noindent Here
\begin{equation}                                                            
v^{\lambda \mu}_{A B} = \frac{1}{2} \left ( f^\lambda_A f^\mu_B - f^\mu_A f^\lambda_B \right ) \sqrt{g}, ~~~ V^{\lambda \mu}_{A B} = \frac{1}{2} \epsilon^{\lambda \mu \nu \rho} f_{\nu A} f_{\rho B}, ~~~ v \circ R \equiv \frac{1}{2} v_{AB} R^{AB},
\end{equation}

\noindent $T_\lambda$ is the translation along the edge $\lambda$ to the next site, $T_\lambda f (x^\lambda ) = f (x^\lambda + 1)$, the overlining in $\overOm$, $\overT$, ... means the Hermitian conjugation.

In (\ref{S-discr}), we have used the exact SO(10) representation for the discrete GR (Regge) action or separately discretized the Cartan-Weyl form from (\ref{S Imm full}). As for the constraint with the Lagrange multipliers $\Lambda$ (the term which violates the local SO(10) symmetry), the property of the vanishing of the horizontal-horizontal block in $\omega$ is not preserved when passing from the generator of the connection matrix to the matrix itself. And, strictly speaking, the equivalence of action (\ref{S-discr}) to the action of GR is obtained for the connection that differs little from unity.

Meanwhile, the kinetic term arising in the limit of continuous time is proportional to ${\rm tr} ( A^{0 \lambda} \overOm_\lambda \dotOm_\lambda ) $, where $A^{\lambda \mu} = v^{\lambda \mu} + V^{\lambda \mu} / \gamma_{\rm F}$, a combination of the direct and dual area bivectors \cite{kha-spectrum}. If $\Omega = \exp \omega$ is close to unity, then, provided that the horizontal-horizontal block of $\omega$ disappears, the kinetic term is proportional to ${\rm tr} ( A^{0 \lambda} \omega_\lambda \dotom_\lambda ) $, where the horizontal-vertical and vertical-horizontal blocks of $\omega$ work. This means that the area spectrum is singular, $\propto 1 / \varepsilon$ on the classical background with such $\omega$ with a scale $\varepsilon$. Thus, an analysis of the classical background with large $\omega$ is of interest.

In the present paper we consider the discrete first order Faddeev action for the large connection matrix and strongly varying vector fields. We find that the equations of motion for connection can be satisfied with the help of a simultaneous redefinition of both the vector fields and connection using the same finite rotation matrices. A periodic cell is supposed to exist such that the variations of the variables from cell to cell are small in order that the continuum limit would exist on large scales. We confine ourselves to a simple ansatz in which connection in only one direction is large. In section \ref{equations} we analyze the equations of motion for connection and show that some properties of the first order Faddeev gravity and natural requirements allow to extend the solution and the local SO(10) violating term to the large $\omega$ region practically uniquely. In section \ref{second} we find the resulting second order action in the leading order as we approach the continuum limit. The action is the sum of the related copies of the Faddeev action. In section \ref{GR} this is reduced to the GR action.

\section{Equations for connection}\label{equations}

Consider the variation of the action with respect to the connection. The action of the operator $\left ( \Omega^C_{\mu A} \partial / \partial \Omega^{CB}_\mu - (A \leftrightarrow B) \right )$ (to remove the implicit orthogonality condition for $\Omega_\mu$ which is added to the action after multiplying by certain Lagrange multipliers; there is no summation over $\mu$!) on the action $S^{\rm discr}$ without the local SO(10) violating term gives
\begin{eqnarray}\label{var-S-wrt-Omega}                                     
\sum_\lambda \left [ \frac{v^{\lambda \mu} R_{\lambda \mu} + \overR_{\lambda \mu} v^{\lambda \mu} }{\cos \alpha_{\lambda \mu}} - T_\lambda \left ( \Omega_\lambda \frac{v^{\lambda \mu} \overR_{\lambda \mu} + R_{\lambda \mu} v^{\lambda \mu} }{\cos \alpha_{\lambda \mu}} \overOm_\lambda  \right ) \right. \nonumber \\ \left. + \frac{1}{\gamma_{\rm F}} \frac{V^{\lambda \mu} R_{\lambda \mu} + \overR_{\lambda \mu} V^{\lambda \mu} }{\cos \alpha^*_{\lambda \mu} } - \frac{1}{\gamma_{\rm F}} T_\lambda \left ( \Omega_\lambda \frac{V^{\lambda \mu} \overR_{\lambda \mu} + R_{\lambda \mu} V^{\lambda \mu} }{\cos \alpha^*_{\lambda \mu}} \overOm_\lambda  \right ) \right ], \nonumber \\ \alpha_{\lambda \mu} = \arcsin \frac{v^{\lambda \mu} \circ R_{\lambda \mu}(\Omega ) }{\sqrt{v^{\lambda \mu} \circ v^{\lambda \mu} } }, ~~~ \alpha^*_{\lambda \mu} = \arcsin \frac{V^{\lambda \mu} \circ R_{\lambda \mu}(\Omega ) }{\sqrt{V^{\lambda \mu} \circ V^{\lambda \mu} } }.
\end{eqnarray}

For the regular case of the weakly varying fields, $\omega$ ($\Omega = \exp \omega$) has the first order of smallness with respect to the finite differences of $f$ at the neighboring sites, and $r$ ($R = \exp r$) - the second,
\begin{equation}                                                            
\omega = O(\delta ), ~~~ r = O(\delta^2), ~~~ \delta_\lambda = 1 - \overT_\lambda
\end{equation}

\noindent (as confirmed by further calculation), which gives a finite contribution to the action in the continuum limit, when $\delta \to \partial$. Expression (\ref{var-S-wrt-Omega}) has the order $O(\delta )$ of interest to us, and in this order we set $R = 1, \cos \alpha = 1$.

The action of $\left ( \Omega^C_{\mu A} \partial / \partial \Omega^{CB}_\mu - (A \leftrightarrow B) \right )$ on the local SO(10) violating term in (\ref{S-discr}) gives
\begin{equation}\label{SO10violate}                                         
\sum_{\nu, \lambda} \Lambda^\mu_{\nu \lambda} \left \{ \overOm_\mu [f^\nu, f^\lambda ] + [f^\nu, f^\lambda ] \Omega_\mu \right \}_{AB} ~~~ ( [f^\nu , f^\lambda]_{AB} \equiv f^\nu_A f^\lambda_B - f^\lambda_A f^\nu_B ).
\end{equation}

\noindent Equate the sum of the expressions (\ref{var-S-wrt-Omega}) and (\ref{SO10violate}) to zero. Acting on each of the indices $A, B$ of the resulting expression by the {\it horizontal} projector $\Pi_{||} = 1 - \Pi$ or evaluating for each $\mu$ the trace of its products with each of the six independent bivectors $[f^\nu , f^\lambda]$, we find that $\Lambda = O (\delta )$. Having found $\Lambda^\mu_{\nu \lambda}$, we can project the equation over one of the indices $A, B$ vertically (with the help of $\Pi$) and find that the contribution of (\ref{SO10violate}) is $O(\delta^2 )$ and can be disregarded. As a result, the equation for $\omega$ in the order $O(\delta )$ takes the form
\begin{equation}\label{w-equ}                                              
\sum_\lambda \Pi T_\lambda \left \{ \delta_\lambda A^{\lambda \mu} + [\omega_\lambda , A^{\lambda \mu}] \right \} = 0 ~~~ (A^{\lambda \mu} \equiv v^{\lambda \mu} + \frac{1}{\gamma_{\rm F}} V^{\lambda \mu} )
\end{equation}

\noindent with the solution
\begin{equation}\label{w-solution}                                         
\omega_\lambda = [f^\mu , \Pi \delta_\lambda f_\mu ].
\end{equation}

We note the following properties of the equations under consideration for $\Omega$.
\begin{itemize}
\item[(i)] To find $\Omega_\lambda$, we should vary over $\Omega_\mu$, $\mu \neq \lambda$.
\item[(ii)] $\Omega_\lambda$ is determined by the dependence of $f$ on $x^\lambda$.
\end{itemize}
We probe this using an ansatz in which $\Omega_\lambda$ is large only at one value of $\lambda$, say 3,
\begin{eqnarray}\label{Omega3}                                             
\Omega_3 (x) = [1 + \omega_3 (x) ] U(x), ~~~ x^3 = \dots , -2, 0, 2, \dots \nonumber  \\
\Omega_3 (x) = [1 + \omega_3 (x) ] \overU (x), ~~~ x^3 = \dots , -1, 1, 3, \dots .
\end{eqnarray}

\noindent To be exact, $\exp \omega = 1 + \omega + \omega^2 / 2 + O ( \delta^3 )$ is required, but symmetric $\omega^2$ does not contribute in the desired order $O ( \delta^2 )$ in what follows. The resulting (approximately) periodic cell with respect to $x^3$ consists of two elementary cells, and with respect to $x^\lambda$, $\lambda \neq 3$ - of one. The order of magnitude $O (\delta )$ is ascribed to $\delta_\lambda U$ and to the variations of $f^\lambda_A (x )$ under the shifts of $x$ by any period. Roughly, equations (\ref{Omega3}) correspond to the additional rotation by $\overU |_{x^3 = 2 n}$ or, within the required accuracy, by $\overU |_{x^3 = 2 n + 1}$ in the 4-cubes at $2n < x^3 < 2n + 1$ (or by $U |_{x^3 = 2 n - 1}$ or $U |_{x^3 = 2 n}$ in the 4-cubes at $2n - 1 < x^3 < 2n$).

Then we can perform the following. \\
1) Using (ii), $\Omega_\lambda$ at $\lambda \neq 3$ are already known, see equation (\ref{w-solution}), and the local SO(10) violating constraints on them are the same as in (\ref{S-discr}). \\
2) Using (i) and 1) for the above constraints, we vary the action with respect to $\Omega_\mu$ at $\mu \neq 3$, find $\Omega_3$ and restore the local SO(10) violating constraint on $\Omega_3$. \\
3) Using (i) and 2) for the above constraint, we vary the action with respect to $\Omega_3$, substitute $\Omega_\lambda$ at $\lambda \neq 3$ from 1) and thus perform the consistency check, which indeed, as it turns out, takes place.

The point 1) here is already fulfilled, and we consider the expression (\ref{var-S-wrt-Omega}) where now still $\cos \alpha = 1$ in the order $O(\delta )$, but, generally speaking, $R - 1 = O(\delta )$ and should be taken into account so that this expression reads
\begin{equation}\label{var-S-wrt-Omega-alpha-0}                            
\sum_\lambda \left \{ A^{\lambda \mu} R_{\lambda \mu} + \overR_{\lambda \mu} A^{\lambda \mu} - T_\lambda \left [ \Omega_\lambda \left ( A^{\lambda \mu} \overR_{\lambda \mu} + R_{\lambda \mu} A^{\lambda \mu} \right ) \overOm_\lambda \right ] \right \}.
\end{equation}

\noindent First consider $\mu \neq 3$ (point 2)). At the points $x^3 = \dots , -1, 1, 3, \dots$, with taking into account (\ref{Omega3}), this expression takes the form
\begin{eqnarray}                                                           
-2 \sum_{\lambda \neq 3} T_\lambda \left \{ \delta_\lambda A^{\lambda \mu} (f ) + [\omega_\lambda , A^{\lambda \mu} (f )] \right \}  \nonumber \\ -2 T_3 \left \{ A^{3 \mu}(U f ) - \overT_3 A^{3 \mu}(f ) + [\omega_3 , A^{3 \mu} (U f )] \right \}
\nonumber \\
+ [A^{3 \mu} (f ), r_{3 \mu}] + T_3 [A^{3 \mu} (U f ), U r_{3 \mu} \overU ].
\end{eqnarray}

\noindent Here $A^{\lambda \mu} (U f )$ means $A^{\lambda \mu}$ composed of $U f$ instead of $f$; $A^{\lambda \mu} (U f ) = U A^{\lambda \mu} (f ) \overU$. Introduce the following field $h^\lambda_A (x )$ which, unlike $f^\lambda_A (x )$, is supposed to have small variation from site to site.
\begin{eqnarray}                                                           
h^\lambda (x ) = U (x ) f^\lambda (x ), ~~~ x^3 = \dots , -2, 0, 2, \dots \nonumber  \\
h^\lambda (x ) = f^\lambda (x ), ~~~ x^3 = \dots , -1, 1, 3, \dots .
\end{eqnarray}

\noindent This allows us to write the expression in a homogeneous manner as
\begin{eqnarray}\label{var-S-Omega-x-1}                                    
-2 \sum_{\lambda } T_\lambda \left \{ \delta_\lambda A^{\lambda \mu} (h ) + [\omega_\lambda , A^{\lambda \mu} (h )] \right \}
+ [A^{3 \mu} (h ), r_{3 \mu}] + T_3 [A^{3 \mu} (h ), U r_{3 \mu} \overU ].
\end{eqnarray}

\noindent Similarly, at the points $x^3 = \dots , -2, 0, 2, \dots$, expression (\ref{var-S-wrt-Omega-alpha-0}) takes the form
\begin{eqnarray}\label{var-S-Omega-x-0}                                    
-2 \sum_{\lambda } T_\lambda \left \{ \delta_\lambda A^{\lambda \mu} (l ) + [\omega_\lambda , A^{\lambda \mu} (l )] \right \}
+ [A^{3 \mu} (l ), r_{3 \mu}] + T_3 [A^{3 \mu} (l ), \overU r_{3 \mu} U ],
\end{eqnarray}

\noindent where one else "almost smooth" field $l^\lambda_A (x )$ is introduced,
\begin{eqnarray}                                                           
l^\lambda (x ) = f^\lambda (x ), ~~~ x^3 = \dots , -2, 0, 2, \dots \nonumber  \\
l^\lambda (x ) = \overU (x ) f^\lambda (x ), ~~~ x^3 = \dots , -1, 1, 3, \dots
\end{eqnarray}

\noindent (its variations $\delta_\lambda l$ tend to zero in the continuum limit). We note that
\begin{equation}                                                           
h^\lambda (x ) = U (x ) l^\lambda (x ).
\end{equation}

\noindent Looking ahead, the expressions for $r_{3 \mu}$ look as (\ref{r-h-x}). The contribution of $r_{3 \mu}$ (the last two terms in (\ref{var-S-Omega-x-1}), (\ref{var-S-Omega-x-0})) can be omitted for any one of the following two reasons. \\
a) The values $r_{3 \mu}$ have the form $\Pi_{||} \dots \Pi_{||} + \Pi \dots \Pi$, and the last two terms in (\ref{var-S-Omega-x-1}), (\ref{var-S-Omega-x-0}) are horizontal in both ten-vector indices; subsequent projecting by $\Pi$ in order to get rid off the Lagrange multipliers $\Lambda$ cancels these terms too. \\
b) The last two terms in (\ref{var-S-Omega-x-1}), (\ref{var-S-Omega-x-0}) cancel each other in the required order $O(\delta )$ since
\begin{eqnarray}                                                           
T_3 (U r_{3 \mu} \overU ) = - r_{3 \mu} + O(\delta^2 ), ~~~ x^3 = \dots , -1, 1, 3, \dots , \nonumber  \\
T_3 (\overU r_{3 \mu} U ) = - r_{3 \mu} + O(\delta^2 ), ~~~ x^3 = \dots , -2, 0, 2, \dots .
\end{eqnarray}

As before, the contribution from the local SO(10) violating term and the dependence on the Lagrange multipliers $\Lambda$ (\ref{SO10violate}) is canceled by the action of the vertical projector $\Pi (f )$. The argument $f$ shows that it is a functional (in fact, a function) of the fields $f^\lambda_A (x )$. At the odd values of $x^3$, this argument is substituted by $h$, and at the even values - by $l$. The equations for $\omega$ take the already appeared form
\begin{eqnarray}                                                           
\sum_\lambda \Pi (h ) T_\lambda \left \{ \delta_\lambda A^{\lambda \mu} (h ) + [\omega_\lambda , A^{\lambda \mu} (h )] \right \} = 0, ~~~ x^3 = \dots , -1, 1, 3, \dots , \nonumber  \\ \label{var-S-Omega-mu-Pi} \sum_\lambda \Pi (l ) T_\lambda \left \{ \delta_\lambda A^{\lambda \mu} (l ) + [\omega_\lambda , A^{\lambda \mu} (l )] \right \} = 0, ~~~ x^3 = \dots , -2, 0, 2, \dots ,
\end{eqnarray}

\noindent the difference only in the fields $h$ and $l$ instead of $f$ at different points. Accordingly, denoting the functional of the solution (\ref{w-solution}) by $\omega_\lambda (f )$, we can write the solution as
\begin{eqnarray}                                                           
\omega_3 = \omega_3 (h ), ~~~ \omega_\lambda = \omega_\lambda (l ), \lambda \neq 3, ~~~ x^3 = \dots , -2, 0, 2, \dots , \nonumber  \\
\omega_3 = \omega_3 (l ), ~~~ \omega_\lambda = \omega_\lambda (h ), \lambda \neq 3, ~~~ x^3 = \dots , -1, 1, 3, \dots .
\end{eqnarray}

It remains to determine the local SO(10) violating constraint for $\Omega_3$ and to check that the equation obtained by variation with respect to $\Omega_3$ is also satisfied. The constraint on $\omega_3$ according to the solution found is similar to that in (\ref{S-discr}); for example, at $x^3 = \dots , -2, 0, 2, \dots$ we have
\begin{equation}                                                           
(1 + \omega_3)^{AB} (h^\lambda_A h^\mu_B - h^\mu_A h^\lambda_B) = 0.
\end{equation}

\noindent Passing to $\Omega_3$, we obtain
\begin{eqnarray}\label{SO10violate-Omega3}                                 
\sum_{\lambda , \mu} \Lambda^3_{\lambda \mu} \Omega_3^{AB} (h^\lambda_A l^\mu_B - h^\mu_A l^\lambda_B)
\end{eqnarray}

\noindent for the corresponding constraint term in the action and a similar equation for odd values of $x^3$, which differs from it by interchange $h \leftrightarrow l$.

The result of applying the operator $\left ( \Omega^C_{3 A} \partial / \partial \Omega^{CB}_3 - (A \leftrightarrow B) \right )$ to the action, equation (\ref{var-S-wrt-Omega-alpha-0}) for $\mu = 3$, for even $x^3$ is written as
\begin{eqnarray}\label{var-S-Omega3}                                       
\sum_{\lambda } \left \{ -2 T_\lambda \left \{ \delta_\lambda A^{\lambda 3} (l ) + [\omega_\lambda (l ) , A^{\lambda 3} (l )] \right \} + [A^{\lambda 3} (l ), r_{\lambda 3}] + T_\lambda [A^{\lambda 3} (l ), r_{\lambda 3} ] \right \}.
\end{eqnarray}

\noindent Again, the contribution of $r_{\lambda 3}$ (the last two terms in braces) can be disregarded, since the values $r_{\lambda 3}$ have the form $\Pi_{||} \dots \Pi_{||} + \Pi \dots \Pi$, and these terms are horizontal in both ten-vector indices; subsequent projecting by $\Pi$ in order to get rid off the Lagrange multipliers $\Lambda$ cancels these terms too. As for these Lagrange multipliers, applying the operator $\left ( \Omega^C_{3 A} \partial / \partial \Omega^{CB}_3 - (A \leftrightarrow B) \right )$ to the term (\ref{SO10violate-Omega3}), we get for their contribution
\begin{equation}\label{SO10violate-vary-Omega3}                            
\sum_{\lambda, \mu} \Lambda^3_{\lambda, \mu} \left \{ 2 [l^\lambda , l^\mu ] + [[l^\lambda , l^\mu ], \overU \omega_3 (h ) U] \right \}.
\end{equation}

\noindent The sum of the expressions (\ref{var-S-Omega3}) and (\ref{SO10violate-vary-Omega3}) is equated to zero. As earlier, projecting this horizontally from both sides, we find that $\Lambda = O (\delta )$; then, projecting by $\Pi (l )$, we make the contribution of the local SO(10) violating term to be $O(\delta^2 )$ and disregard it. The result is the equation
\begin{equation}                                                           
\sum_\lambda \Pi (l ) T_\lambda \left \{ \delta_\lambda A^{\lambda 3} (l ) + [\omega_\lambda , A^{\lambda 3} (l )] \right \} = 0
\end{equation}

\noindent similar to the earlier found (\ref{var-S-Omega-mu-Pi}) and satisfied by the found solution $\omega_\lambda (l )$. The consistency check for odd $x^3$ is similar.

\section{Second order action}\label{second}

We substitute the solutions $\Omega (f )$ found into the expressions $R^{AB}_{\lambda \mu} (\Omega )$. We write out the generator $r$ of $R = \exp r$, which in the considered order $O (\delta^2 )$ is simply the antisymmetric part of $R$,
\begin{equation}                                                           
r = \frac{1}{2} ( R - \overR ) = \frac{1}{2} ( \tilde{R} - \overtilR ),
\end{equation}

\noindent where $\tilde{R}$ 
differs from $R$ by that we substitute into the expression $R (\Omega )$ for $\Omega_\lambda = \exp \omega_\lambda , \lambda \neq 3$ (or $\Omega_3 = (\exp \omega_3 ) U$ ) the value $1 + \omega_\lambda$ (or $(1 + \omega_3) U$ ) disregarding the $\omega^2$ term (and it is not necessarily an orthogonal matrix because of this). If $\lambda , \mu \neq 3$, then
\begin{eqnarray}                                                           
\tilde{R}_{\lambda \mu} = [1 - \omega_\lambda (f )] [1 - ( \overT_\lambda \omega_\mu (f ) )] [1 + ( \overT_\mu \omega_\lambda (f ) )] [1 + \omega_\mu (f )], \nonumber \\
r_{\lambda \mu} = r_{\lambda \mu} (\omega ( f ) ), ~~~ r_{\lambda \mu} (\omega ) \equiv \delta_\lambda \omega_\mu - \delta_\mu \omega_\lambda + [ \omega_\lambda , \omega_\mu ],
\end{eqnarray}

\noindent where $f = h$ for odd $x^3$ or $f = l$ for even $x^3$.

For the case when one of the indices is 3, we consider, say, for even $x^3$
\begin{eqnarray}                                                           
\tilde{R}_{13} = U [1 - \omega_1 (l )] ( \overT_1 \overU ) [1 - ( \overT_1 \omega_3 (h ) )] [1 + ( \overT_3 \omega_1 (h ) )] [1 + \omega_3 (h )].
\end{eqnarray}

\noindent We use
\begin{equation}\label{U-omega-l-U=omega-h}                                
U \omega_1 (l ) \overU = U \omega_1 (\overU h ) \overU = \omega_1 (h ) - \Pi (h ) U (\delta_1 \overU ) \Pi_{||} (h ) + \Pi_{||} (h ) ( \delta_1 U ) \overU \Pi (h ),
\end{equation}

\noindent where, in turn, we have used $U \Pi (l ) \overU = \Pi (U l ) = \Pi (h )$. The result for even $x^3$ is
\begin{eqnarray}                                                           
& & U r_{13} \overU = \frac{1}{2} [ (\delta_1 U ) \overU - U (\delta_1 \overU ) ] - \Pi_{||} (h ) (\delta_1 U ) \overU \Pi (h ) + \Pi (h ) U (\delta_1 \overU ) \Pi_{||} (h ) \nonumber \\ & & - \frac{1}{2} [ \Pi_{||} (h ) U (\delta_1 \overU ) \Pi_{||} (h ) + \Pi (h ) U (\delta_1 \overU ) \Pi (h ) , \omega_1 (h ) ] + \frac{1}{2} [U \omega_1 (l ) \overU , U (\delta_1 \overU ) ] \nonumber \\ & & + r_{13} ( \omega (h ) ), ~~~ r_{13} (\omega ) \equiv \delta_1 \omega_3 - \delta_3 \omega_1 + [ \omega_1 , \omega_3 ].
\end{eqnarray}

\noindent For odd $x^3$, the interchange $U \leftrightarrow \overU$ and $h \leftrightarrow l$ should be made. In particular, the $O(\delta )$ parts can be singled out which are used in the equations for connection above,
\begin{eqnarray}\label{r-h-x}                                              
& & \hspace{-5mm} r_{13} = \Pi_{||} (h ) U ( \delta_1 \overU ) \Pi_{||} (h ) + \Pi (h ) U ( \delta_1 \overU ) \Pi (h ) + O(\delta^2 ), ~ x^3 = \dots , -1, 1, 3, \dots, \nonumber \\
& & \hspace{-5mm} r_{13} = \Pi_{||} (l ) \overU ( \delta_1 U ) ] \Pi_{||} (l ) + \Pi (l ) \overU ( \delta_1 U ) ] \Pi (l ) + O(\delta^2 ), ~ x^3 = \dots , -2, 0, 2, \dots.
\end{eqnarray}

We see that in any case there is a standard term $r_{\lambda \mu} (\omega ( f ) )$, $f = h$ or $l$ which arises when connection is excluded from the equations of motion in the regular case of small $\Omega - 1$ ($U = 1$). In addition, there are some additional terms, if $\lambda$ or $\mu$ is 3. Let us write out their contributions to the action, $-{\rm tr} A^{13} R_{13}$,
\begin{eqnarray}\label{A13R13}                                             
{\rm tr} A^{13} (h ) [U ( \delta_1 \overU ) - U \omega_1 (l ) ( \delta_1 \overU )], ~~~ x^3 = \dots , -2, 0, 2, \dots , \nonumber  \\ {\rm tr} A^{13} (l ) [\overU ( \delta_1 U ) - \overU \omega_1 (h ) ( \delta_1 U )], ~~~ x^3 = \dots , -1, 1, 3, \dots.
\end{eqnarray}

\noindent Let us express these both in terms of $h$. At even $x^3$, we reduce $U \omega_1 (l ) \overU $ to $\omega_1 (h )$ plus some form linear in $U ( \delta_1 \overU )$ (\ref{U-omega-l-U=omega-h}). The latter form gives a zero contribution to (\ref{A13R13}), since either $A^{13}$ is canceled by the projector $\Pi$, or the trace of the product of the symmetric matrix $U ( \delta_1 \overU ) \Pi U ( \delta_1 \overU )$ with antisymmetric $A^{13}$ is taken. At odd $x^3$, we use $A^{13} (l ) = \overU A^{13} (h ) U$. The contributions (\ref{A13R13}) take the form
\begin{eqnarray}\label{A13R13-h}                                           
& & {\rm tr} A^{13} (h ) [U ( \delta_1 \overU ) - \omega_1 (h ) U ( \delta_1 \overU )] \equiv F(x^3), ~~~ x^3 = \dots , -2, 0, 2, \dots , \nonumber  \\ & & {\rm tr} A^{13} (h ) [ ( \delta_1 U ) \overU - \omega_1 (h ) ( \delta_1 U ) \overU ] = \nonumber \\ & & = {\rm tr} A^{13} (h ) [ - U ( \delta_1 \overU ) + \omega_1 (h ) U ( \delta_1 \overU )] = - F(x^3), ~~~ x^3 = \dots , -1, 1, 3, \dots.
\end{eqnarray}

\noindent At odd $x^3$, we replaced the expression $ ( \delta_1 U ) \overU $ by $ - U ( \delta_1 \overU )$ in both terms in square brackets. In the second term, we write $ ( \delta_1 U ) \overU = - U ( \delta_1 \overU ) + m$, $m = O( \delta^2 )$ at $\delta \to \partial$ according to the product differentiation rule. In the first term, $O( \delta^2 )$ can not be neglected, but it is easy to see that $m$ is symmetric and does not contribute.

Thus, the additional contributions to the action from sites with coordinates $x^3$ and $x^3 + 1$ are close in value and opposite in sign. Let us have an integer number $N$ of periodic cells along $x^3$. The sum of the contributions (\ref{A13R13-h}) over $x^3$ can be transformed into a similar alternating sum, but already of finite differences,
\begin{equation}\label{sum-F}                                              
\sum^{2N}_{n = 1} ( - 1 )^n F (n ) = \frac{1}{2} \sum^{2N}_{n = 2} ( - 1 )^n \delta_3 F (n ) + \frac{1}{2} F (2N ) - \frac{1}{2} F (1 ).
\end{equation}

\noindent Repeating this procedure already with respect to the latter sum, we find that only boundary terms ($F (2N ) / 2 - F (1 ) / 2$) contribute in the continuum limit. That is, the contribution (\ref{A13R13-h}) effectively can be considered as a full derivative in this limit, $\delta_3 F / 2 \to \partial_3 F / 2$. Moreover, further iterations show that, up to certain boundary terms, the sum is close to zero, even to an arbitrarily high order $O( \delta^n )$ (for a sufficiently high order of differentiability of the function).

The resulting bulk contributions are
\begin{eqnarray}\label{AR}                                                 
A^{\lambda \mu}_{A B} (l ) r^{A B}_{\lambda \mu} ( \omega (l ) ) ~ (\lambda , \mu \neq 3 ), ~ A^{\lambda 3}_{A B} (h ) r^{A B}_{\lambda 3} ( \omega (h ) ), ~~ x^3 = \dots , -2, 0, 2, \dots , \nonumber  \\ A^{\lambda \mu}_{A B} (h ) r^{A B}_{\lambda \mu} ( \omega (h ) ) ~ (\lambda , \mu \neq 3 ), ~ A^{\lambda 3}_{A B} (l ) r^{A B}_{\lambda 3} ( \omega (l ) ), ~~ x^3 = \dots , -1, 1, 3, \dots.
\end{eqnarray}

\noindent It is noteworthy that different contributions that depend on $h$ (as well as those that depend on $l$) are taken at $x^3$ of different parity. However, as one can paraphrase the remark just made after (\ref{sum-F}), the sum over even points tends to that over odd points in the continuum limit (and even is close to that to within an arbitrarily high order $O(\delta^n )$ depending on the order of differentiability of the interpolating continuum fields $h$, $l$). In total, the contributions (\ref{AR}) look as the result of excluding connection from the continuum first order action (\ref{S Imm full}) in the finite difference approximation. Correspondingly, at half the points, these formulas reproduce half the Faddeev action for the fields $h$, in the remaining half - for the fields $l$.

\section{Equivalence to general relativity}\label{GR}

Thus, the considered system is described in the continuum limit by the sum of the Faddeev type actions,
\begin{eqnarray}\label{Faddeev+Imm}                                        
& & \frac{1}{2} \int \Pi^{AB} (h ) \left [ (h^\lambda_{A, \lambda} h^\mu_{B, \mu} - h^\lambda_{A, \mu} h^\mu_{B, \lambda}) \sqrt {g} - \frac{1}{\gamma_{\rm F}} \epsilon^{\lambda \mu \nu \rho} h_{\lambda A, \mu} h_{\nu B, \rho} \right ] \d^4 x \nonumber \\ & & + \frac{1}{2} \int \Pi^{AB} (l ) \left [ (l^\lambda_{A, \lambda} l^\mu_{B, \mu} - l^\lambda_{A, \mu} l^\mu_{B, \lambda}) \sqrt {g} - \frac{1}{\gamma_{\rm F}} \epsilon^{\lambda \mu \nu \rho} l_{\lambda A, \mu} l_{\nu B, \rho} \right ] \d^4 x \nonumber \\ & & + \int M_{\lambda \mu} (h^\lambda h^\mu - l^\lambda l^\mu ) \sqrt{g } \d^4 x.
\end{eqnarray}

\noindent The last term is a constraint on the fields $h$, $l$ taken into account with the help of a symmetric Lagrange multiplier matrix $M_{\lambda \mu}$. This constraint ensures existence of an SO(10) matrix $U$ such that $h = U l$ and simultaneously unambiguity of the metric $g^{\lambda \mu} = h^\lambda h^\mu = l^\lambda l^\mu$.

Applying the operator $\Pi_{AB} (h ) \delta / ( \sqrt{g} \delta h^\lambda_B ) $, we cancel the constraint term and find the vertical equations of motion,
\begin{equation}                                                           
\hspace{-0mm} b^\mu_{\mu A} T^\nu_{\lambda \nu} + b^\mu_{\lambda A} T^\nu_{\nu \mu} + b^\mu_{\nu A} T^\nu_{\mu \lambda} + \frac{\epsilon^{\tau \mu \nu \rho}}{2 \gamma_{\rm F} \sqrt{g}} (g_{\lambda \sigma} g_{\kappa \tau} - g_{\lambda \tau} g_{\kappa \sigma}) b^{\kappa}_{\rho A} T^\sigma_{\mu \nu} = 0.
\end{equation}

\noindent This is temporarily a functional of $h$, $T^\lambda_{\mu \nu} = h^\lambda_A (h^A_{\mu , \nu} - h^A_{\nu , \mu} )$, $b^\lambda_{\mu A} = \Pi_{AB} (h ) h^{\lambda B}_{, \mu}$. The modification associated with the parity violating term, especially since one more parameter is added ($\gamma_{\rm F}$), seems to be not qualitatively crucial, and these equations, like those that do not have this term (\ref{vertical}), still give $T^\lambda_{\mu \nu} = 0$ almost everywhere in the infinite-dimensional configuration superspace. This reduces the $h$-dependent part of the action (\ref{Faddeev+Imm}) to half the Hilbert-Einstein action. Similarly, the $l$-dependent part of the action gives the remaining half. Hence the Hilbert-Einstein action follows.

\section{ Conclusion}

Thus, the discrete first order representation of the Faddeev gravity works also if the connection is large. This allows us to have a reasonable area spectrum and, at the same time, classically, GR on a large scale. (Some example of consistent area spectrum of our work \cite{kha-spectrum} corresponds to $U = \exp ( \pi [ n_3 , p_3 ] ) \exp ( \pi [ n_1 , p_1 ] )$ where $n_\lambda$ are mutually orthogonal unit vectors in the horizontal subspace, $p_\lambda$ are arbitrary unit vectors in the vertical subspace.)

In the second order Faddeev gravity, the direct substitution of the piecewise constant fields with non-negligible discontinuities into the Lagrangian density leads to products of step functions and delta functions. The resulting action is finite and, in general, depends on the intermediate regularization of these products. The first order formalism can be considered as such a regulator.

In our transition to the second order formalism, the curvature supports are located on the boundary between the regions with the fields $l$ and $h = U l$ (if $l$ is taken as independent), and their contributions are {\it mixed} functionals of both $h$ and $l$. Despite this, the sum of the contributions of the neighboring sites has been reduced to the sum of the functions of the {\it separate} arguments $h$ and $l$. Because of the local gauge SO(10) non-invariance of the Faddeev action, a dependence on $U$ exists and disappears only indirectly, by excluding vector fields on going to GR.

In these calculations, an important point was specifying the (discrete) local gauge violating constraint thus far known for the connection matrices close to unity. This could be done in the course of extending the solutions for the connection, using some natural assumptions about the continuation of the theory to the strongly varying fields (items (i), (ii) after eq (\ref{w-solution})). Our result for the constraint term like (\ref{SO10violate-Omega3}) can be written for the general coordinateless simplicial formulation in terms of the edge components as
\begin{equation}\label{SO10violate-simplicial}                             
\Lambda^{\sigma^3 \sigma^1_1 \sigma^1_2 } \Omega^{AB}_{\sigma^3} [h_{\sigma^1_1 A} (\sigma^4_1 ) l_{\sigma^1_2 B} (\sigma^4_1 ) - h_{\sigma^1_2 A} (\sigma^4_1 ) l_{\sigma^1_1 B} (\sigma^4_1 ) ],
\end{equation}

\noindent now for the covariant edge components $h^A_{\sigma^1} \equiv h^A_\lambda \Delta x^\lambda_{\sigma^1}$, etc.; $\Delta x^\lambda_{\sigma^1}$ is the world vector of the edge $\sigma^1$. Here the 3-simplex $\sigma^3$ is shared by the 4-simplices $\sigma^4_1$ and $\sigma^4_2$, and $\Omega_{\sigma^3}$ acts from $\sigma^4_2$ to $\sigma^4_1$. It is assumed that there is some periodic cell, and the values of $f$ on some sets of analogous 4-simplices coincide with the values of $h$ and $l$ which are some fields smooth in the continuum limit and such that $h (\sigma^4_1 ) = f (\sigma^4_1 )$, $l (\sigma^4_2 ) = f (\sigma^4_2 )$. However, in (\ref{SO10violate-simplicial}) $h$ and $l$ are taken in the {\it same} 4-simplex, here $\sigma^4_1$ (this just violates the local SO(10)). The pairs of its edges $(\sigma^1_1, \sigma^1_2)$ form full set of six independent bivectors.

This allows, in particular, to examine the case when $\Omega_\lambda$ for all $\lambda$ can significantly differ from unity. The new is that then the curvature matrices have $O(1)$ parts like $\overOm_1 \overOm_2 \Omega_1 \Omega_2$, leading to infinity in the continuum limit. Therefore, $\Omega_\lambda$ can not be freely chosen, and the corresponding conditions on $\Omega$ should be imposed.

\section*{Acknowledgments}

The present work was supported by the Ministry of Education and Science of the Russian Federation

\end{document}